\documentclass[sigconf,screen]{acmart}

\usepackage{microtype}
\usepackage{graphicx}
\usepackage{subfigure}
\usepackage{booktabs} 
\usepackage{mathtools}
\usepackage{multirow}
\usepackage{xcolor}
\usepackage{colortbl}
\usepackage{bm}
\usepackage{array}
\usepackage{fancyhdr}
\usepackage{hyperref}
\usepackage{enumitem}
\usepackage{makecell} 
\usepackage[skip=2pt]{caption}
\usepackage[ruled,linesnumbered]{algorithm2e}
\AtBeginDocument{%
  }

\newcommand{\headernodot}[1]{\vspace*{1mm}\noindent{\textbf{#1}}}
\newcommand{\header}[1]{\headernodot{#1.}}
\author{Liu Yang}
\affiliation{%
  \institution{Shandong University}
  \city{Qingdao}
  \country{China}
}
\email{yangliushirry@gmail.com}

\author{Zhaochun Ren}
\affiliation{%
  \institution{Leiden University}
  \city{Leiden}
  \country{The Netherlands}
}
\email{z.ren@liacs.leidenuniv.nl}

\author{Ziqi Zhao}
\affiliation{%
  \institution{Shandong University}
  \city{Qingdao}
  \country{China}
}
\email{ziqizhao.work@gmail.com}

\author{Pengjie Ren}
\affiliation{%
  \institution{Shandong University}
  \city{Qingdao}
  \country{China}
}
\email{jay.ren@outlook.com}

\author{Zhumin Chen}
\affiliation{%
  \institution{Shandong University}
  \city{Qingdao}
  \country{China}
}
\email{chenzhumin@sdu.edu.cn}

\author{Xinyi Li}
\affiliation{%
  \institution{Baidu Inc.}
  \city{Beijing}
  \country{China}
}
\email{lixinyimichael@163.com}

\author{Zhiming Peng}
\affiliation{%
  \institution{Baidu Inc.}
  \city{Beijing}
  \country{China}
}
\email{pengzhiming01@baidu.com}

\author{Daiting Shi}
\affiliation{%
  \institution{Baidu Inc.}
  \city{Beijing}
  \country{China}
}
\email{shidaiting01@baidu.com}

\author{Maarten de Rijke}
\affiliation{%
  \institution{University of Amsterdam}
  \city{Amsterdam}
  \country{The Netherlands}
}
\email{m.derijke@uva.nl}

\author{Xin Xin}
\authornote{Corresponding author.}
\affiliation{%
  \institution{Shandong University}
  \city{Qingdao}
  \country{China}
}
\email{xinxin@sdu.edu.cn}

\renewcommand{\shortauthors}{Liu Yang et al.}



\begin{document}

\title[Curriculum Approximate Unlearning for Session-based Recommendation]{Curriculum Approximate Unlearning for \\ Session-based Recommendation}

\begin{abstract}

Approximate unlearning aims to eliminate the influence of specific training samples from a recommender system without the need for model retraining.
Gradient ascent (GA) is a prominent method for approximate unlearning but applying it to session-based recommendation presents two primary challenges. 
First, a naive application of GA often results in a significant degradation of recommendation performance.
Second, existing studies fail to account for the sequential ordering of unlearning samples when processing multiple  simultaneously unlearning requests, leading to sub-optimal efficacy and recommendation performance.
To address these challenges, we introduce CAU, a \textbf{c}urriculum \textbf{a}pproximate \textbf{u}nlearning framework tailored to session-based recommendation. 
CAU executes unlearning by applying a GA term to the target unlearning samples. 
To mitigate performance degradation (the first challenge), CAU treats unlearning as a multi-objective optimization problem. By integrating the GA term with retaining terms for preserving performance, the framework identifies a Pareto-Optimal solution. This approach ensures robust unlearning with minimal impact on  recommendation performance.
To resolve the second challenge regarding sample ordering, CAU adopts a curriculum-based sequence to process unlearning batches. 
Motivated by the principle of progressing from ``easy'' to ``hard'' samples, we introduce two metrics to quantify unlearning difficulty: \emph{gradient difficulty} and \emph{embedding difficulty}. We further propose two selection strategies, hard-sampling and soft-sampling, to prioritize unlearning samples on these difficulty scores. 
Extensive experiments on three benchmark datasets, alongside comprehensive ablation studies, demonstrate that each component of CAU contributes significantly to its superior performance and unlearning efficacy.
\end{abstract}

\maketitle

\section{Introduction}
Session-based recommendation has shown remarkable effectiveness in predicting users' future interests based on their historical interactions.
While these systems excel at memorizing sequential patterns, the ability to eliminate the influence of specific training samples, a.k.a. \emph{recommendation unlearning} (RU), has emerged as equally crucial.
Several data protection regulations, such as the EU General Data Protection Regulation (GDPR), the California Consumer Privacy Act (CCPA), and the China Personal Information Protection Law (PIPL), mandate individuals’ rights to have their personal data removed from trained recommendation models~\cite{mantelero2013eu,baik2020data,schelter-2023-forget}.
Beyond this legal perspective, RU serves additional practical purposes, such as removing outdated user preferences~\cite{liu2022forgetting,matuszyk2015forgetting} or eliminating the impact of noisy training samples~\cite{shaik2023exploring,schelter-2024-snapcase}.

\header{Recommendation unlearning} Previous studies on RU can be categorized into exact recommendation unlearning and approximate recommendation unlearning \cite{li2024survey}. 

\emph{Exact recommendation unlearning} eradicates unlearning data impact by removing samples and retraining the model~\cite{nguyen2022survey}. To improve efficiency, exact RU methods typically partition data and train sub-models on each shard~\cite{chen2022recommendation,chen2022graph,sisa}, then aggregate results. However, when numerous unlearning requests occur concurrently, efficiency deteriorates substantially as unlearning data spans multiple shards, requiring retraining of several sub-models or the entire model. Moreover, due to item correlations, exact RU methods may fail to achieve true unlearning~\cite{xin2024effectiveness}, as removed interactions can still be inferred from correlated interactions through collaborative signals.
 
\emph{Approximate recommendation unlearning} offers a more efficient alternative by enabling rapid forgetting and lowering overall computational costs. 
It efficiently removes the influence of specific training samples from a recommendation model without retraining the entire model or sub-models from scratch~\cite{li2024survey}.
Gradient ascent (GA) is a representative method that applies opposite gradients to unlearning samples, increasing the loss function to mitigate their influence from a parametric perspective. \citet{yao2024largelanguagemodelunlearning} proposed to use GA-based methods for unlearning in large language models. 

\header{Challenges} 
Despite its success, directly applying GA to conduct unlearning for session-based recommendation still poses non-trivial challenges:
\begin{enumerate}[label=(\arabic*),leftmargin=*]
\item Naive application of GA for session-based recommendation can severely degrade performance. Because GA's objective function is the inverse of gradient descent, it may diverge without specific strategies, causing training failure. Furthermore, due to sequential correlations, performing GA on one session can negatively impact the accuracy of others. Consequently, balancing unlearning efficacy with maintaining performance via GA remains a significant challenge.

\item Existing research overlooks the sequence of unlearning samples when processing simultaneous requests. While curriculum learning effectively improves training by progressing from easy to hard samples, the impact of unlearning orders remains underexplored for session-based recommendation. Intuitively, sequential correlations between interactions suggest that the ordering of unlearning instances significantly influences the final outcome.
\end{enumerate}

\header{Our proposed method} We propose \emph{curriculum approximate unlearning} (CAU), a framework for session-based recommendation that uses gradient ascent (GA) to remove data influence. To address the first challenge and maintain recommendation quality, CAU formulates unlearning as a multi-objective optimization problem, balancing the GA term with performance-retention objectives. By identifying a Pareto-optimal solution, CAU achieves effective unlearning with minimal performance loss.

To address the second challenge, CAU employs a curriculum-based sequence to process unlearning batches. We quantify difficulty using two metrics: \emph{gradient unlearning difficulty}, which measures the consistency between GA and retaining gradients, and \emph{embedding unlearning difficulty}, which captures the semantic proximity between session states and target items. Based on these scores, CAU implements two strategies: \emph{hard-sampling}, a deterministic approach strictly following difficulty rankings, and \emph{soft-sampling}, a stochastic approach that prioritizes easier samples early on before gradually shifting toward harder ones.

To verify CAU's effectiveness, we implement it on three representative session-based recommendation models, viz.\ GRU4Rec, SASRec, and BERT4Rec. We conduct extensive experiments on three benchmark datasets. Our experimental results show CAU's effectiveness.

\header{Contributions} To summarize, our main contributions in this paper are:
\begin{itemize}[leftmargin=*] 
    \item We propose \emph{CAU}, the first curriculum-based multi-task optimization framework for approximate unlearning in session-based recommendation, using Pareto-optimization to balance unlearning efficacy with performance retention. 
    \item We introduce \emph{gradient} and \emph{embedding unlearning difficulty} metrics to quantify the complexity of removing specific samples in session-based contexts. 
    \item We design two \emph{curriculum learning strategies} (hard-sampling and soft-sampling) that optimize the unlearning process by progressing from easy to difficult samples. 
    \item \emph{Extensive experiments} across three benchmark datasets and three state-of-the-art models demonstrate that CAU achieves superior unlearning efficiency and effectiveness while maintaining high recommendation accuracy. 
\end{itemize}

\section{Related Work}

\subsection{Session-based recommendation}
Session-based recommendation aims to predict a user’s next action based on the current session.
Early methods rely on session-based nearest-neighbor techniques, which recommend items frequently co-occurring with the current session~\cite{hariri2012context,hariri2015adapting,bell2007scalable}. 
These methods are simple and interpretable but struggle to capture complex sequential patterns.

With the rise of deep learning, RNN-based models~\cite{donkers2017sequential,hidasi2018recurrent,mu2018survey} were proposed to model session sequences, achieving better performance by capturing temporal dependencies. 
Later works explored self-attention and Transformer-based architectures to model long-range dependencies and complex item interactions within sessions~\cite{kang2018self,sun2019bert4rec,xu2019graph,fang2021session}. 
Recent research further incorporates graph neural networks (GNNs) to represent sessions as graphs~\cite{wu2019session,wang2019neural,qiu2019rethinking}, capturing intricate item transition patterns and higher-order connectivity.
Beyond model architecture, there are efforts on improving training and personalization strategies, including contrastive learning~\cite{liu2021contrastive,xie2022contrastive}, and meta-learning~\cite{song2021cbml,pan2022multimodal}.
These approaches aim to better exploit session context, mitigate data sparsity, and improve generalization to unseen sessions.

\subsection{Machine unlearning}
Machine unlearning aims to remove the influence of certain training data from a given model. 
In traditional tasks like classification, previous studies have explored concepts related to machine unlearning, such as data deletion and selective forgetting~\cite{datadeletion,guo2023certifieddataremovalmachine,izzo2021approximatedatadeletionmachine,shibata2021learning}.
And current machine unlearning methods generally fall into two main categories: exact unlearning and approximate unlearning. 

Exact unlearning methods typically rely on reversible training procedures or special model structures~\cite{cao2015towards,sisa,chen2022graph}, enabling the model to reconstruct its state as if the target samples were never seen. 
In contrast, approximate unlearning methods focus on efficiently reducing the residual impact of unlearning data, often by fine-tuning the model parameters with negative gradients~\cite{golatkar2020eternal,sekhari2021remember} or applying targeted weight perturbations~\cite{graves2021amnesiac,guo2019certified}. 
While approximate methods cannot guarantee perfect removal, they offer significant advantages in scalability and computational efficiency~\cite{nguyen2022survey}.

While several contemporary studies have adapted multi-task optimization (MTO) to LLM unlearning~\cite{jin-etal-2025-unlearning,pan2025multiobjectivelargelanguagemodel}, it is crucial to note that recommendation tasks are fundamentally distinct from general LLM-based tasks. Specifically, recommendation systems primarily rely on collaborative signals and user-item interaction graphs, which exhibit higher data sparsity compared to the dense semantic spaces of LLMs. Consequently, directly migrating MTO frameworks from LLMs to recommendation unlearning without considering these domain-specific characteristics may lead to suboptimal performance and the loss of critical collaborative information.

\subsection{Recommendation unlearning}
Unlearning in the recommendation scenario tends to attract more research attention. 
Recommendation unlearning can not only help to protect user privacy but also improve recommendation models through eliminating the effect of noisy data and misleading information~\cite{shaik2023exploring,lubitzsch-2025-towards}.
\citet{liu2022forgetting} and~\citet{xu2023netflix} proposed to use fine-tuning and the alternative least square algorithm for unlearning acceleration. \citet{chen2022recommendation} and~\citet{li2022making} extended the ideas of the SISA algorithm for collaborative filtering. 
And~\citet{hu2025exact} applied a similar idea to LLM-based recommendation unlearning.
\citet{xin2024effectiveness} proposed an exact unlearning framework for session-based recommendation.
Additionally, there are other techniques like matrix correction~\cite{liu2023recommendation}, influence functions~\cite{zhang2024recommendation}, and label-flipping~\cite{alshehri2023forgetting}.
However, none of the existing approximate unlearning methods is tailored for session-based recommendation. 
Recently, \citet{10.1145/3705328.3748092} introduced interaction-level unlearning difficulty metrics for collaborative filtering to guide extra deletion but does not leverage difficulty to optimize the unlearning process.
In this work, we propose two methods to estimate the difficulty of unlearning samples and introduce two corresponding sampling strategies to implement a curriculum learning scheme, thereby further improving the effectiveness of the unlearning process.

\section{Task Definition}
In this section, we describe session-based recommendation and then formulate the task of unlearning in this context.

\subsection{Session-based recommendation}
Let $V$ be the set of items, $S=\{v_1, \ldots, v_t, \ldots, v_n\}$ be the user-item interaction sequence in a session, where $v_t \in V$ is the item interacted at time step $t$ in the session.  
The dataset $D$ of user-item interaction sessions is defined as $D=\{S_1,\ldots,S_{|D|}\}$. 
Given a user-item interaction sequence at time step $t$, the learning objective of session-based recommendation is to predict the next item $v_{t+1}$ that the user is likely to interact with, which can be formalized as:
\begin{equation}
\mathbf{p}_{t+1}= \mathcal{M}_\theta(v|S, D),
\end{equation}
where $\mathbf{p}_{t+1}$ denotes the predicted probability distribution over candidate items at time step $t+1$, $\mathcal{M}_\theta$ represents the recommendation model parameterized by $\theta$, e.g., GRU4Rec \cite{hidasi2015session} or SASRec \cite{kang2018self}.
Typically, the recommendation model $\mathcal{M}_\theta$ can be trained by optimizing a loss function $\mathcal{L}_{rec}$ based on the predicted probability distribution $\mathbf{p}_{t+1}$, such as the cross-entropy loss or the pair-wise ranking loss. 
The training process can be formalized as:
\begin{equation}
\theta^{rec}=\arg\min_{\theta} \mathcal{L}_{rec}(\theta^{inital};D),
\end{equation}
where $\theta^{initial}$ denotes the initial model parameters, and $\theta^{rec}$ represents the optimized parameters obtained by minimizing the recommendation loss $\mathcal{L}_{rec}$ over the dataset $D$.

\begin{figure*}[htbp]
  \centering
  \includegraphics[width=0.8\textwidth]{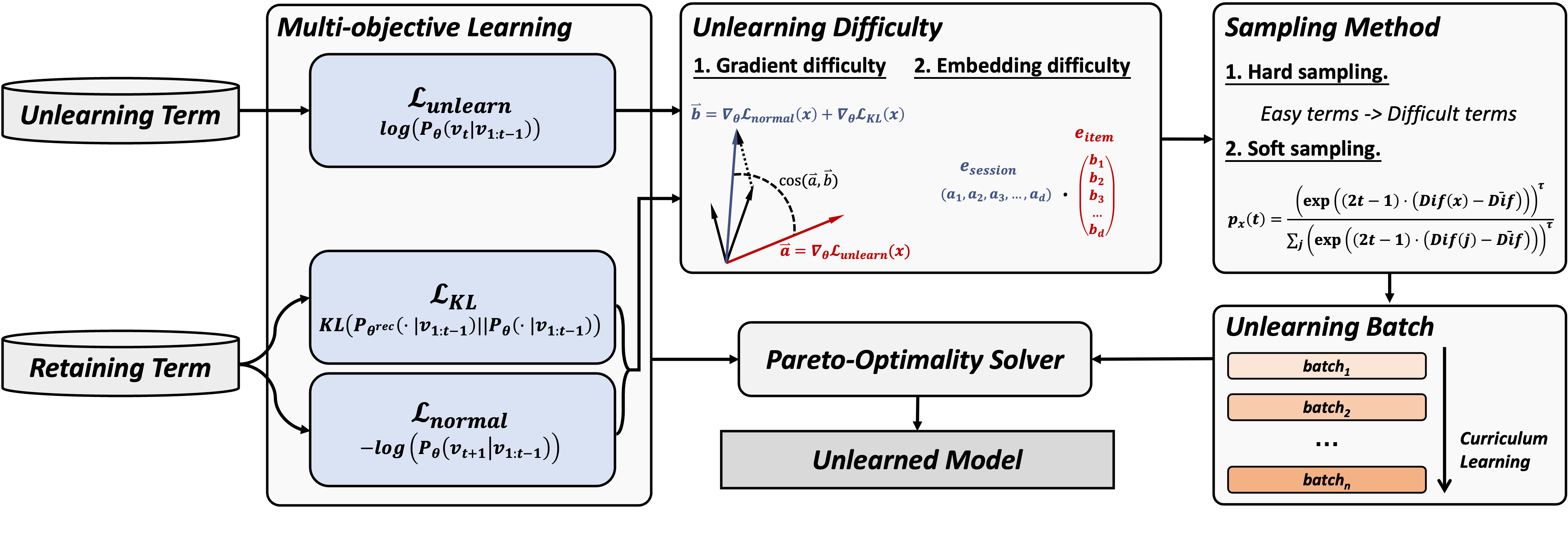}
  \caption{Overview of the proposed CAU framework. CAU defines an unlearning objective $\mathcal{L}_{unlearn}$ and remaining terms $\mathcal{L}_{normal}$ and $\mathcal{L}_{KL}$. Then CAU introduces two metrics to measure the unlearning difficulty, including gradient difficulty and embedding difficulty. Two sampling methods are proposed to sample batches from easy to hard, according to the unlearning difficulty metrics. Finally, a Pareto-optimality solver is used to perform multi-objective learning to obtain the unlearned model.}
  \label{fig:model_overview}
\end{figure*}

\subsection{Unlearning}
To address privacy concerns or enhance recommendation utility, recommender systems need to support unlearning requests to remove the impact of specific samples.
For example, users might request to eliminate traces of interactions with sensitive items to preserve their privacy.
Existing methods include \emph{exact recommendation unlearning} and \emph{approximate recommendation unlearning}. 

\emph{Exact recommendation unlearning} is aimed at completely eradicating the impact of the data to be forgotten as if they never occurred in the training process. 
Let $D_f \subset D$ denotes the data to be unlearned, the naive approach to achieve exact unlearning is to retrain the model from scratch on the dataset without the unlearning data, i.e., $D-D_f$.
This process can be defined as:
\begin{equation}
\theta^{exa}=\arg\min_{\theta} \mathcal{L}_{rec}(\theta^{inital};D-D_f),
\end{equation}
where $\theta^{exa}$ represents the parameters of exact unlearned model obtained by minimizing the recommendation loss $\mathcal{L}_{rec}$ over $D-D_f$.

In contrast to exact unlearning, \emph{approximate recommendation unlearning} offers a computationally more efficient alternative that aims to emulate the effect of exact unlearning, while preserving model utility and avoiding retraining.
Instead of retraining the model from scratch, approximate unlearning optimizes the model parameters directly in response to unlearning requests.
Common methods for approximate unlearning include gradient ascent on unlearning samples to reverse their influence \cite{graves2021amnesiac,sekhari2021remember} or filtering the model’s outputs \cite{baumhauer2022machine}.
Approximate recommendation unlearning can be defined as:
\begin{equation}
\theta^{app}=\arg\min_{\theta} \mathcal{L}_{app}(\theta^{rec};D_f),
\end{equation}
where $\theta^{app}$ denotes the updated model parameters obtained by optimizing a specially designed approximate unlearning loss $\mathcal{L}_{app}$, 
starting from the previous recommendation model parameters $\theta^{rec}$.

\section{Method}
In this section, we detail the operational mechanism of CAU, which primarily consists of multi-objective learning and curriculum-based sample selection.

The framework is illustrated in Figure \ref{fig:model_overview}. 
To maintain recommendation performance after unlearning, CAU formulates the task as a multi-objective learning problem.
CAU incorporates a learning objective that focuses on preserving recommendation performance, alongside the unlearning objective for the unlearning data $D_f$. 
We then consider the model's gradient and embedding space and introduce two metrics to measure unlearning difficulty, respectively.
Based on these difficulty metrics, CAU proposes two sampling methods, \textit{hard-sampling} and \textit{soft-sampling}, to construct unlearning batches, which facilitates a curriculum strategy to perform unlearning from easy to hard. 
Finally, a Pareto-optimality solver is used to achieve multi-objective optimization to obtain the unlearned model parameters.

\subsection{Multi-objective unlearning}
While effective unlearning is important, maintaining recommendation performance is equally crucial.
This necessitates a careful balance between the unlearning objective and the recommendation utility, so we formulate  approximate unlearning as a multi-task optimization problem. 
The key idea is to jointly optimize the recommendation utility and unlearning effectiveness through training a composite objective function with three task-specific losses: (i)~an unlearning loss $\mathcal{L}_{unlearn}$ that encourages forgetting specific samples, (ii)~a retention loss $\mathcal{L}_{normal}$ to preserve recommendation performance, and (iii)~a consistency loss $\mathcal{L}_{KL}$ that regularizes deviations from the previous model. 

Next, we introduce the three types of loss function in detail and describe how they are weighted during optimization.

\textbf{Unlearning loss.} 
Unlearning can be regarded as the inverse process of forward learning, so we apply gradient ascent to explicitly remove the influence of the unlearning sample, which intuitively corresponds to adding back the gradient of the target data that was previously subtracted. 
For the target item $v_t$ to be unlearned, the unlearning loss is formulated as:
\begin{equation}
\mathcal{L}_{unlearn} = \log(P_{\theta}(v_t|v_{1:t-1})).
\end{equation}
This formulation effectively pushes the model away from favoring the unlearning item, by reducing its predicted probability in the output distribution.

\textbf{Normal loss.} 
To ensure that the unlearned model retains its general predictive capability, we include a standard cross-entropy loss based on the next item $v_{t+1}$, given the prefix session $\{v_1, \ldots, v_{t-1}\}$:
\begin{equation}
\mathcal{L}_{normal} = -\log(P_{\theta}(v_{t+1}|v_{1:t-1})).
\end{equation}
Since unlearning is performed to remove an item from a session, it is essential to ensure that the model maintains continuity on the remaining items in the session.
$\mathcal{L}_{normal}$ guides the model to remain functional in its primary task, i.e., predicting plausible future items based on user behavior, despite forgetting specific items.

\textbf{KL loss.} 
While removing unwanted information, CAU aims to minimize catastrophic drift from the previous model to better maintain the model's recommendation utility. 
Therefore, we include a KL divergence loss to align the output distribution of the unlearned model $P_{\theta}$ with the previous model $P_{\theta^{rec}}$:
\begin{equation}
\mathcal{L}_{KL} = KL(P_{\theta^{rec}}(\cdot|v_{1:t-1})||P_{\theta}(\cdot|v_{1:t-1})).
\end{equation}
This ensures that for remaining samples, the behavior of the new model does not deviate significantly from the previous model, improving stability and recommendation consistency.

Finally, the loss function for multi-objective unlearning is  formulated as follows:
\begin{equation}
\label{eq:L_app}
\mathcal{L}_{app} = \alpha_1\mathcal{L}_{unlearn} + \alpha_2\mathcal{L}_{normal} + \alpha_3\mathcal{L}_{KL}.
\end{equation}

\subsection{Pareto-optimality solver}
A crucial challenge in multi-task optimization is balancing the importance of each objective. 
Manually tuning the weights $\alpha_1$, $\alpha_2$, and $\alpha_3$ is suboptimal and sensitive to data distribution and model scale. 
Furthermore, maintaining fixed hyperparameters throughout the entire training process can hinder the model's ability to adapt to changes and negatively impact performance. 
To effectively solve this problem, we employ a Pareto-optimality solver, which can reliably find a stable solution that balances the trade-offs between conflicting objectives. 
This approach ensures that the model simultaneously achieves effective unlearning while maintaining overall recommendation quality.
Given a model parameterized by $\theta$ and a series of learning objective functions $\mathcal{L}_{unlearn},\mathcal{L}_{normal},\mathcal{L}_{KL}$, 
Pareto-optimality is a state in which it is impossible to enhance one objective without compromising another.
Formally, Pareto-optimality can be defined as:
\begin{definition}
A solution $\theta$ \textbf{dominates} $\bar\theta$ if $\mathcal{L}_k(\theta) \leq\mathcal{L}_k(\bar\theta)$ for all learning objective functions $\mathcal{L}_k$.
\end{definition}

\begin{definition}
A solution $\theta^{*}$ is called \textbf{Pareto-optimal} if there exists no solution $\theta$ that dominates $\theta^{*}$.
\end{definition}

\noindent
Given the training objective $\mathcal{L}_{app}$ in Eq.~(\ref{eq:L_app}), CAU aims to find $\alpha_i$ so that the solution is Pareto-optimal.
Therefore, we employ the multiple gradient descent algorithm (MGDA) to find a solution that balances the competing objectives. 
MGDA ~\cite{sener2018multi} seeks a convex combination of gradients from all objectives such that their weighted sum forms a common descent direction.

Specifically, at each training step, we compute the gradients of each individual task loss w.r.t.\ the model parameters $\theta$:
\begin{equation}
\mathbf{g}_1 = \nabla_{\theta} \mathcal{L}_{unlearn}, \quad
\mathbf{g}_2 = \nabla_{\theta} \mathcal{L}_{normal}, \quad
\mathbf{g}_3 = \nabla_{\theta} \mathcal{L}_{KL}.
\end{equation}
Formally, MGDA uses the Karush-Kuhn-Tucker (KKT)
conditions \cite{bertsekas1997nonlinear} to describe constraints on  $\alpha_1,\alpha_2,\alpha_3$:
\begin{equation}\label{eq:mgda}
\min_{\alpha_1,\alpha_2,\alpha_3} \left\| \sum_{i=1}^3 \alpha_i \mathbf{g}_i \right\|^2_2
\quad \text{subject to} \quad \sum_{i=1}^3 \alpha_i = 1,\ \alpha_i \geq 0.
\end{equation}
Employing MGDA solver to Eq.~(\ref{eq:mgda}),
CAU obtains optimal task weights $\alpha_i$ for each learning objective, ensuring that the training progress is in a direction which improves all tasks without favoring one while sacrificing others. 

After computing $\alpha_i$ for all tasks, we use the loss weighted by $\alpha_i$ to update the model parameters, thus identifying an optimization direction that preserves the performance of recommendation while improving the effectiveness of unlearning.

\subsection{Curriculum unlearning}
To further enhance effectiveness of approximate unlearning when handling multiple requests simultaneously, we propose a curriculum-based unlearning strategy that prioritizes forgetting easier samples first, progressively moving toward harder ones. 
The underlying assumption is that removing low-impact easier samples earlier helps stabilize training and facilitates optimization. 
The curriculum unlearning framework comprises two key components: (i)~difficulty estimation and (ii)~sample scheduling.

\subsubsection{Difficulty estimation}
In this study, we consider two perspectives for measuring unlearning difficulty, including gradient unlearning difficulty and embedding unlearning difficulty.

\textbf{Gradient unlearning difficulty.}
This metric evaluates how difficult it is to update the model in a direction that jointly improves (or at least does not degrade) all individual tasks. 
Since we adopt MGDA for multi-objective optimization, large conflicts between the gradients of individual objectives would hinder the search for a Pareto-optimal solution.

To quantify this, we compute the cosine similarity between the unlearning gradient and the sum of the gradients of the normal and KL losses. 
Specifically, for an unlearning sample $x$, the gradient unlearning difficulty is defined as:
\begin{equation}
\label{eq:dif_g}
\mbox{}\hspace*{-2mm}
\mathit{Dif}_g(x) \!=\! -\cos\left( \nabla_{\theta} \mathcal{L}_{unlearn}(x),\ \nabla_{\theta} \mathcal{L}_{normal}(x) + \nabla_{\theta} \mathcal{L}_{KL}(x)\right)\!.
\hspace*{-1mm}\mbox{}
\end{equation}
A higher cosine similarity implies better alignment among the objectives, meaning the sample is easier to unlearn. 
Therefore, we sort the samples by increasing cosine similarity and prioritize those with smaller gradient conflicts in the beginning.

Gradient unlearning difficulty measures the difficulty from an external perspective, i.e., measuring the degree of conflicts between unlearning and retaining.

\textbf{Embedding unlearning difficulty.}
This metric is based on the intuition that samples which the model strongly favors are harder to be forgotten. 
We compute the dot product between the session state representation and the embedding of the item to be forgotten.
Given a session state representation $\mathbf{e}_{session}$ and the embedding of the target item $x$ to be forgotten $\mathbf{e}_{x}$, CAU computes:
\begin{equation}
\label{eq:dif_e}
\mathit{Dif}_e(x) = \langle \mathbf{e}_{session},\ \mathbf{e}_{x} \rangle.
\end{equation}
Here, $\mathbf{e}_{session}$ is the encoded state representation of the session prefix (e.g., from an RNN or self-attention network), and $\mathbf{e}_{x}$ is the embedding of the target item $x$ to be forgotten. $\langle \rangle$ denotes dot product. 
A higher $\mathit{Dif}_e$ implies the model is more confident in predicting the unlearning target $x$ , making it harder to be removed. 
Hence, we treat samples with lower $\mathit{Dif}_e$ as easier ones and schedule them earlier in the unlearning process.

\begin{algorithm}
\SetAlgoLined
\KwIn{unlearning data $D_f$, training epochs $E$, recommender $\mathcal{M}$ with parameter $\theta^{rec}$}
\KwOut{unlearned model $\mathcal{M}_u$ with parameter $\theta^{app}$}

$\mathcal{M}_{u} \gets$ initialize unlearned model with model $\mathcal{M}$\;
\For{$epoch \gets 1$ \KwTo $E$}{
    calculate $\mathit{Dif}_g$ or $\mathit{Dif}_e$ according to Eq.~(\ref{eq:dif_g}) or Eq.~(\ref{eq:dif_e})\;
    $D_{f\text{-sorted}} \gets \text{sort } D_f \text{ ascendingly } \text{by $\mathit{Dif}_g$ or $\mathit{Dif}_e$}$\; 
    sample $batch$ $B$ sequentially from $D_{f\text{-sorted}}$\;
    \ForEach{batch $B$}{
        $\alpha_1,\alpha_2,\alpha_3 \gets$ calculate weights by Eq.~(\ref{eq:mgda})\; 
        $\mathcal{L}_{app} \gets$ calculate loss using Eq.~(\ref{eq:L_app})\;
        $\theta^{app} \gets$ update model parameter according to $\mathcal{L}_{app}$\;
    }
}
\Return{$\mathcal{M}_{u}$ \textnormal{with} $\theta^{app}$ }
\caption{CAU training with hard-sampling}
\label{algorithm:hard_sampling}
\end{algorithm}

Embedding unlearning difficulty measures the difficulty from an internal perspective, i.e., whether the recommender itself tends to recommend the target unlearning item $x$. 

In practice, both difficulty metrics can be used to guide the following curriculum sampling for unlearning.

\subsubsection{Sample scheduling}
CAU introduces two strategies to achieve the unlearning curriculum: hard-sampling and soft-sampling.

\textbf{Hard-sampling.} 
In the hard-sampling strategy, the training process for each epoch deterministically adheres to the easy-to-hard curriculum schedule. 
Specifically, all samples in the unlearning dataset are firstly ranked according to their difficulty scores.
The training then proceeds sequentially following the  order, beginning with samples deemed easier to be forgotten and gradually advancing towards more challenging samples.


For example, when employing the gradient unlearning difficulty metric, the initial phase of each epoch concentrates on samples with high gradient alignment (i.e., lower $\mathit{Dif}_g$), indicating that their gradients align well and thus are easier to be optimized. 
As training progresses, more challenging samples with higher $\mathit{Dif}_g$ and more conflicting gradients are incrementally introduced into the training batches. 
Such curriculum progression reduces abrupt gradient conflicts and helps maintain a stable optimization trajectory.

The detailed procedure of the hard-sampling strategy is formalized and presented in Algorithm \ref{algorithm:hard_sampling}.

\textbf{Soft-sampling.} 
To introduce stochasticity into the unlearning process and mitigate the risk of overfitting to a fixed curriculum, CAU proposes a soft-sampling mechanism where samples are drawn stochastically at each step based on their difficulty scores.
Unlike hard-sampling which deterministically follows a strict curriculum order, soft-sampling assigns a dynamic probability distribution over the samples, enabling more flexible and robust training.

Formally, the sampling probability for each sample $x$ at time step $t$ is defined as follows:
\begin{equation}
\label{eq:soft}
p_x(t) = \frac{(\exp( (2t-1) \cdot (\mathit{Dif}(x) - \overline{\mathit{Dif}} )))^{\tau}}{\sum_j (\exp((2t-1) \cdot (\mathit{Dif}(j) - \overline{\mathit{Dif}})))^{\tau}},
\end{equation}
where $\mathit{Dif}$ denotes the difficulty scores (i.e., either $\mathit{Dif}_g$ or $\mathit{Dif}_e$), and $\overline{\mathit{Dif}}$ is the mean difficulty score of all unlearning samples. 
The time variable $t$ is normalized to the interval $[0,1]$ based on the current training progress, calculated as $t=\mathit{current\_step}/\mathit{total\_step}$. $\tau$ is the temperature coefficient. 

\begin{algorithm}[htb]
\SetAlgoLined
\KwIn{unlearning data $D_f$, total training steps $T$, recommender $\mathcal{M}$ with parameter $\theta^{rec}$}
\KwOut{unlearned model $\mathcal{M}_u$ with parameter $\theta^{app}$}

$\mathcal{M}_u \gets$ initialize unlearned model with model $\mathcal{M}$\;


\For{$step \gets 1$ \KwTo $T$}{
    $t \gets step / T$\;
    \ForEach{sample $x$ in $D_f$}{
    calculate $\mathit{Dif}$ according to Eq.~(\ref{eq:dif_g}) or Eq.~(\ref{eq:dif_e})\;
        $p_x(t) \gets $ calculate the sampling probability according to Eq.~(\ref{eq:soft})\; 
    }
    
    $B \gets$ sample a batch from $D_f$ according to $p_x(t)$\;
    
    $\alpha_1,\alpha_2,\alpha_3 \gets$ calculate weights by Eq.~(\ref{eq:mgda})\; 
        $\mathcal{L}_{app} \gets$ calculate loss using Eq.~(\ref{eq:L_app})\;
        $\theta^{app} \gets$ update model parameter according to $\mathcal{L}_{app}$\;
}
\Return{$\mathcal{M}_{u}$ \textnormal{with} $\theta^{app}$ }
\caption{CAU training with soft-sampling}
\label{algorithm:soft_sampling}
\end{algorithm}



Soft-sampling ensures that at the early stages of training ($t \approx 0$), the exponent $(2t - 1)$ is close to $-1$, which inversely weights the difficulty score, making easier samples with lower difficulty scores more likely to be selected. 
As training progresses ($t \approx 1$), the exponent $(2t - 1)$ approaches $+1$, shifting the sampling preference toward harder samples. 
This step-dependent adjustment of the sampling distribution can be viewed as a form of curriculum learning with soft transitions, allowing the model to gradually focus on increasingly difficult samples without abrupt switches. 
Algorithm \ref{algorithm:soft_sampling} illustrates the overall procedure of soft-sampling.

\section{Experiments}

\subsection{Experimental settings}
\subsubsection{Datasets and pre-processing}
We conduct experiments on three public accessible datasets: MovieLens-1M,\footnote{\url{https://grouplens.org/datasets/movielens/1m/}} Amazon Beauty and Games.\footnote{\url{https://jmcauley.ucsd.edu/data/amazon/}} 
These datasets vary in scale, domain, and sparsity.


Following the data pre-processing procedure of \cite{sun2019bert4rec}, we filter out users and items with fewer than 5 interactions and sort user interactions chronologically to form sequences.
For each dataset, we adopt cross-validation to evaluate the performance, and we randomly split sessions into training set, validation set, and test set with a ratio of 8:1:1.

\subsubsection{Evaluation protocols}
The recommendation performance is measured with two metrics: $Recall@k$ and Normalized Discounted Cumulative Gain$@k$
($NDCG@k$). 
$Recall@k$ measures how many ground-truth items are included in the top-$k$ positions of the recommendation list. 
$NDCG@k$ is a rank-sensitive metric that assigns higher weights to top positions in the recommended list.

To evaluate unlearning effectiveness, we adopt the $Hit_u@k$ metric introduced in \cite{xin2024effectiveness}. 
It is used to measure the unlearning effect through assessing whether the unlearned item reappears in the top-$k$ recommendation list generated by the unlearned model, based on the remaining interactions within the session. Lower $Hit_u@k$ denotes better unlearning effectiveness.

Recommendation performance and unlearning effectiveness follow distinct numerical patterns, such as higher $NDCG@k$ and $Recall@k$ indicate better recommendation performance, while lower $Hit_u@k$ signifies better unlearning effectiveness.
We therefore introduce a new metric $U_{\beta}$ to facilitate a comprehensive evaluation of both aspects. 
$U_{\beta}$ integrates $Recall@k$ and $Hit_{u}@k$ as:
\begin{equation}
U_{\beta} = (1 + \beta^2) \cdot \frac{Recall \times (1-Hit_u)}{\beta^2 \times Recall + (1-Hit_u)}.
\end{equation}
The hyperparameter $\beta$ controls the relative emphasis between recommendation performance and unlearning effectiveness. 
This formulation enables a comprehensive evaluation that reflects the trade-off between maintaining high recommendation quality and ensuring effective removal of specific items from the model’s output. 

\subsubsection{Baselines}
Three representative session-based recommendation models are used as the backbone models.
\begin{itemize}[leftmargin=*]
\item \textbf{GRU4Rec} \cite{hidasi2015session} uses gated recurrent units (GRU) to model user interaction sequences which is an RNN-based sequential recommendation model.
\item \textbf{SASRec} \cite{kang2018self} uses the left-to-right unidirectional Transformer \cite{vaswani2017attention} decoder for session-based recommendation.
\item \textbf{BERT4Rec} \cite{sun2019bert4rec} employs  bidirectional self-attention to model interaction sequences.
\end{itemize}

\begin{table*}[htbp]
\centering
\caption{Recommendation performance comparison after unlearning 10\% of data. N is short for NDCG, R is short for Recall. Best results other than Original and Retrain are highlighted in bold.}
\label{table:recommendation_performance}
\resizebox{0.95\textwidth}{!}{
\begin{tabular}{l cccc cccc cccc}
\toprule
Dataset / Method & \multicolumn{4}{c}{\textbf{GRU4Rec}} & \multicolumn{4}{c}{\textbf{SASRec}} & \multicolumn{4}{c}{\textbf{BERT4Rec}} \\
\cmidrule(r){2-5}
\cmidrule(r){6-9}
\cmidrule{10-13}
\textbf{Beauty} & \textbf{N@10} & \textbf{N@20} & \textbf{R@10} & \textbf{R@20} & \textbf{N@10} & \textbf{N@20} & \textbf{R@10} & \textbf{R@20} & \textbf{N@10} & \textbf{N@20} & \textbf{R@10} & \textbf{R@20} \\
\midrule
Original & 0.0317 & 0.0368 & 0.0533 & 0.0739 & 0.0264 & 0.0325 & 0.0513 & 0.0756 & 0.0270 & 0.0339 & 0.0511 & 0.0781 \\
Retrain  & 0.0326 & 0.0375 & 0.0562 & 0.0756 & 0.0286 & 0.0350 & 0.0512 & 0.0770 & 0.0256 & 0.0314 & 0.0496 & 0.0728 \\
SISA     & 0.0107 & 0.0135 & 0.0153 & 0.0262 & 0.0055 & 0.0082 & 0.0120 & 0.0225 & 0.0089 & 0.0179 & 0.0103 & 0.0232 \\
SRU      & 0.0191 & 0.0226 & 0.0332 & 0.0455 & 0.0178 & 0.0223 & 0.0332 & 0.0511 & 0.0175 & 0.0172 & 0.0332 & 0.0409 \\
\textbf{CAU+\textit{gradient+hard}} & 0.0272 & 0.0333 & 0.0484 & 0.0725 & \textbf{0.0271} & \textbf{0.0327} & 0.0527 & 0.0747 & 0.0245 & 0.0307 & 0.0460 & \textbf{0.0710} \\
\textbf{CAU+\textit{gradient+soft}} & 0.0254 & 0.0308 & 0.0485 & 0.0700 & 0.0259 & 0.0320 & 0.0513 & \textbf{0.0758} & \textbf{0.0246} & 0.0302 & \textbf{0.0475} & 0.0695 \\
\textbf{CAU+\textit{embedding+hard}} & \textbf{0.0289} & \textbf{0.0349} & \textbf{0.0529} & \textbf{0.0769} & 0.0267 & 0.0313 & \textbf{0.0536} & 0.0717 & 0.0245 & \textbf{0.0308} & 0.0465 & 0.0697 \\
\textbf{CAU+\textit{embedding+soft}} & 0.0253 & 0.0304 & 0.0483 & 0.0686 & 0.0255 & 0.0308 & 0.0522 & 0.0732 & 0.0240 & 0.0300 & 0.0440 & 0.0678 \\
\midrule
 \textbf{Games} & \textbf{N@10} & \textbf{N@20} & \textbf{R@10} & \textbf{R@20} & \textbf{N@10} & \textbf{N@20} & \textbf{R@10} & \textbf{R@20} & \textbf{N@10} & \textbf{N@20} & \textbf{R@10} & \textbf{R@20} \\
\midrule
Original & 0.0569 & 0.0713 & 0.1049 & 0.1618 & 0.0461 & 0.0590 & 0.1004 & 0.1511 & 0.0565 & 0.0706 & 0.1096 & 0.1660 \\
Retrain  & 0.0593 & 0.0703 & 0.0110 & 0.1538 & 0.0498 & 0.0640 & 0.1041 & 0.1607 & 0.0528 & 0.0666 & 0.1012 & 0.1557 \\
SISA     & 0.0092 & 0.0115 & 0.0170 & 0.0261 & 0.0119 & 0.0149 & 0.0196 & 0.0318 & 0.0172 & 0.0189 & 0.0203 & 0.0515 \\
SRU      & 0.0197 & 0.0226 & 0.0337 & 0.0485 & 0.0266 & 0.0239 & 0.0501 & 0.0569 & 0.0293 & 0.0361 & 0.0557 & 0.0834 \\
\textbf{CAU+\textit{gradient+hard}} & \textbf{0.0553} & \textbf{0.0680} & \textbf{0.1066} & \textbf{0.1572} & 0.0489 & \textbf{0.0640} & 0.0940 & \textbf{0.1544} & 0.0500 & 0.0651 & \textbf{0.1001} & 0.1555 \\
\textbf{CAU+\textit{gradient+soft}} & 0.0523 & 0.0656 & 0.1011 & 0.1544 & 0.0489 & 0.0627 & 0.0960 & 0.1513 & 0.0502 & 0.0645 & 0.0969 & 0.1548 \\
\textbf{CAU+\textit{embedding+hard}} & 0.0552 & 0.0672 & 0.1053 & 0.1534 & 0.0493 & 0.0638 & 0.0967 & 0.1529 & \textbf{0.0508} & \textbf{0.0655} & 0.0997 & \textbf{0.1586} \\
\textbf{CAU+\textit{embedding+soft}} & 0.0534 & 0.0651 & 0.1059 & 0.1524 & \textbf{0.0502} & 0.0633 & \textbf{0.0974} & 0.1522 & 0.0500 & 0.0635 & 0.0963 & 0.1534 \\
\midrule
 \textbf{ML-1M} & \textbf{N@10} & \textbf{N@20} & \textbf{R@10} & \textbf{R@20} & \textbf{N@10} & \textbf{N@20} & \textbf{R@10} & \textbf{R@20} & \textbf{N@10} & \textbf{N@20} & \textbf{R@10} & \textbf{R@20} \\
\midrule
Original & 0.1315 & 0.1521 & 0.2361 & 0.3171 & 0.1243 & 0.1482 & 0.2155 & 0.3314 & 0.1302 & 0.1538 & 0.2315 & 0.3261 \\
Retrain  & 0.1269 & 0.1512 & 0.2241 & 0.3201 & 0.1325 & 0.1590 & 0.2454 & 0.3546 & 0.1293 & 0.1564 & 0.2300 & 0.3377 \\
SISA     & 0.0212 & 0.0343 & 0.0391 & 0.0915 & 0.0185 & 0.0254 & 0.0317 & 0.0592 & 0.0172 & 0.0242 & 0.0203 & 0.0416 \\
SRU      & 0.0747 & 0.0919 & 0.1255 & 0.1938 & 0.0430 & 0.0497 & 0.0660 & 0.0966 & 0.0412 & 0.0506 & 0.0611 & 0.0977 \\
\textbf{CAU+\textit{gradient+hard}} & 0.1280 & 0.1485 & 0.2311 & 0.3096 & 0.1129 & 0.1328 & \textbf{0.2132} & \textbf{0.2928} & 0.1064 & 0.1286 & 0.2010 & 0.2943 \\
\textbf{CAU+\textit{gradient+soft}} & 0.1329 & \textbf{0.1531} & 0.2351 & \textbf{0.3103} & 0.1018 & 0.1206 & 0.1910 & 0.2644 & 0.1080 & 0.1284 & 0.2063 & 0.2825 \\
\textbf{CAU+\textit{embedding+hard}} & \textbf{0.1333} & 0.1514 & \textbf{0.2357} & 0.3075 & \textbf{0.1136} & \textbf{0.1347} & 0.2088 & 0.2918 & 0.1104 & \textbf{0.1309} & 0.2048 & 0.2906 \\
\textbf{CAU+\textit{embedding+soft}} & 0.1318 & 0.1485 & 0.2326 & 0.3000 & 0.1018 & 0.1242 & 0.1926 & 0.2825 & \textbf{0.1156} & 0.1300 & \textbf{0.2075} & \textbf{0.2946} \\
\bottomrule
\end{tabular}
}
\end{table*}

\noindent%
Each model is trained with the following model-agnostic unlearning frameworks:
\begin{itemize}[leftmargin=*]
\item \textbf{Original}: the original model without unlearning.
\item \textbf{Retrain}: retrain the whole model from scratch on the remaining dataset; this is computationally expensive.
\item \textbf{SISA} \cite{sisa}: a fundamental exact unlearning method that randomly splits the data and averages the outputs of the sub-model.
\item \textbf{SRU} \cite{xin2024effectiveness}: a newly proposed method which improves SISA by collaborative sharding and extra deletion.
\item \textbf{CAU}: the proposed method, comprising the following variants:
    \begin{itemize}
    \renewcommand{\labelitemi}{$+$}
    \renewcommand{\labelitemii}{$+$}
        \item \textit{\textbf{gradient}}: use the gradient unlearning difficulty.
        \item \textit{\textbf{embedding}}: use the embedding unlearning difficulty.
        \item \textit{\textbf{hard}}: use the hard-sampling strategy.
        \item \textit{\textbf{soft}}: use the soft-sampling stragety.
    \end{itemize}

\end{itemize}

\subsubsection{Implementation details}

All methods are learned with the Adam optimizer \cite{kingma2014adam}. 
The learning rate is set as $1 e - 3$. 
The model input consists of the most recent 10 interacted items for Beauty, 20 for Games, and 50 for MovieLens-1M, with shorter sequences padded using a special padding token. 
The training batch size is set to 256, while the unlearning batch size is set to 128. The temperature coefficient $\tau$ is 2.
The embedding size is fixed at 64.
For the unlearning process in hard-sampling, the unlearning epoch is set to 200 for SASRec and BERT4Rec, and 100 for GRU4Rec.
And for soft-sampling, the number of unlearning step is adjusted correspondingly. 
Each experiment is repeated 5 times, and the average performance is reported.

\begin{table*}[htbp]
\centering
\caption{Unlearning effectiveness and overall performance comparison of different methods on the three datasets. The U-score is computed by Recall@10 and $Hit_u$@1, with $\beta=10$ for Beauty and $\beta=3$ for Games and ML-1M. Boldface denotes the best score.}
\label{table:unlearning_effectiveness}
\resizebox{0.85\textwidth}{!}{
\begin{tabular}{l ccc ccc ccc}
\toprule
Dataset / Method & \multicolumn{3}{c}{\textbf{GRU4Rec}} & \multicolumn{3}{c}{\textbf{SASRec}} & \multicolumn{3}{c}{\textbf{BERT4Rec}} \\
\cmidrule(r){2-4}
\cmidrule(r){5-7}
\cmidrule{8-10}
\textbf{Beauty} & \textbf{Hit$_u$@1$\downarrow$} & \textbf{Hit$_u$@5$\downarrow$} & \textbf{U-score$\uparrow$} & \textbf{Hit$_u$@1$\downarrow$} & \textbf{Hit$_u$@5$\downarrow$} & \textbf{U-score$\uparrow$} & \textbf{Hit$_u$@1$\downarrow$} & \textbf{Hit$_u$@5$\downarrow$} & \textbf{U-score$\uparrow$} \\
\midrule
Original & 0.6961 & 0.8950 & 0.2904 & 0.6355 & 0.8319 & 0.3437 & 0.2169 & 0.4285 & 0.6857 \\
Retrain  & 0.1470 & 0.2685 & 0.7479 & 0.1520 & 0.2822 & 0.7347 & 0.1408 & 0.2853 & 0.7396 \\
SISA     & 0.0648 & 0.1640 & 0.5857 & 0.0579 & 0.1569 & 0.5330 & 0.0489 & 0.1457 & 0.6273 \\
SRU      & 0.0822 & 0.1918 & 0.7261 & 0.0828 & 0.1968 & 0.7259 & 0.0775 & 0.1740 & 0.7291 \\
\textbf{CAU+\textit{gradient+hard}} & 0.1262 & 0.2007 & 0.7476 & 0.0973 & 0.1821 & 0.7784 & 0.1324 & 0.2648 & 0.7373 \\
\textbf{CAU+\textit{gradient+soft}} & 0.1088 & 0.1681 & 0.7603 & 0.0823 & 0.1647 & 0.7862 & 0.1296 & 0.2564 & \textbf{0.7430} \\
\textbf{CAU+\textit{embedding+hard}} & 0.1125 & 0.2094 & \textbf{0.7675} & 0.0954 & 0.1784 & 0.7817 & 0.1290 & 0.2672 & 0.7410 \\
\textbf{CAU+\textit{embedding+soft}} & 0.1010 & 0.1793 & 0.7655 & 0.0848 & 0.1635 & \textbf{0.7865} & 0.1283 & 0.2632 & 0.7349 \\
\midrule
\textbf{Games} & \textbf{Hit$_u$@1$\downarrow$} & \textbf{Hit$_u$@5$\downarrow$} & \textbf{U-score$\uparrow$} & \textbf{Hit$_u$@1$\downarrow$} & \textbf{Hit$_u$@5$\downarrow$} & \textbf{U-score$\uparrow$} & \textbf{Hit$_u$@1$\downarrow$} & \textbf{Hit$_u$@5$\downarrow$} & \textbf{U-score$\uparrow$} \\
\midrule
Original & 0.4721 & 0.7669 & 0.3762 & 0.4061 & 0.7022 & 0.3982 & 0.2897 & 0.5765 & 0.4588 \\
Retrain  & 0.2113 & 0.4278 & 0.4884 & 0.2109 & 0.4389 & 0.4759 & 0.1875 & 0.4129 & 0.4771 \\
SISA     & 0.0632 & 0.1610 & 0.1463 & 0.0533 & 0.1760 & 0.1654 & 0.0524 & 0.1589 & 0.2255 \\
SRU      & 0.0816 & 0.1912 & 0.2534 & 0.0691 & 0.1799 & 0.3374 & 0.0691 & 0.1620 & 0.3622 \\
\textbf{CAU+\textit{gradient+hard}} & 0.1619 & 0.3196 & 0.4970 & 0.1521 & 0.3353 & 0.4704 & 0.1679 & 0.3596 & 0.4806 \\
\textbf{CAU+\textit{gradient+soft}} & 0.1449 & 0.2765 & 0.4898 & 0.1410 & 0.3221 & 0.4787 & 0.1585 & 0.3507 & 0.4761 \\
\textbf{CAU+\textit{embedding+hard}} & 0.1525 & 0.3161 & 0.4971 & 0.1423 & 0.3255 & 0.4799 & 0.1636 & 0.3562 & \textbf{0.4810} \\
\textbf{CAU+\textit{embedding+soft}} & 0.1457 & 0.2867 & \textbf{0.5006} & 0.1402 & 0.3034 & \textbf{0.4823} & 0.1585 & 0.3477 & 0.4743 \\
\midrule
\textbf{ML-1M} & \textbf{Hit$_u$@1$\downarrow$} & \textbf{Hit$_u$@5$\downarrow$} & \textbf{U-score$\uparrow$} & \textbf{Hit$_u$@1$\downarrow$} & \textbf{Hit$_u$@5$\downarrow$} & \textbf{U-score$\uparrow$} & \textbf{Hit$_u$@1$\downarrow$} & \textbf{Hit$_u$@5$\downarrow$} & \textbf{U-score$\uparrow$} \\
\midrule
Original & 0.3126 & 0.5983 & 0.5771 & 0.2774 & 0.6025 & 0.5850 & 0.2857 & 0.6066 & 0.5844 \\
Retrain  & 0.2733 & 0.5694 & 0.5936 & 0.2650 & 0.5673 & 0.6127 & 0.2464 & 0.5404 & 0.6139 \\
SISA     & 0.0212 & 0.0841 & 0.2876 & 0.0414 & 0.1636 & 0.2443 & 0.0207 & 0.0849 & 0.1708 \\
SRU      & 0.0714 & 0.2059 & 0.5662 & 0.0538 & 0.2215 & 0.4055 & 0.0042 & 0.0798 & 0.3935 \\
\textbf{CAU+\textit{gradient+hard}} & 0.1325 & 0.2987 & 0.6802 & 0.1366 & 0.2899 & \textbf{0.6616} & 0.0932 & 0.1967 & 0.6711 \\
\textbf{CAU+\textit{gradient+soft}} & 0.1201 & 0.2650 & 0.6905 & 0.1139 & 0.2857 & 0.6497 & 0.0642 & 0.1905 & \textbf{0.6913} \\
\textbf{CAU+\textit{embedding+hard}} & 0.1284 & 0.2547 & 0.6864 & 0.1366 & 0.3043 & 0.6573 & 0.1035 & 0.2298 & 0.6702 \\
\textbf{CAU+\textit{embedding+soft}} & 0.0994 & 0.2526 & \textbf{0.6997} & 0.1118 & 0.2692 & 0.6525 & 0.0911 & 0.2257 & 0.6793 \\
\bottomrule
\end{tabular}
}
\end{table*}

\subsection{Recommendation performance}

We compare the recommendation performance results in Table~\ref{table:recommendation_performance}. 
We can observe that due to the removal of 10\% of the training data, the recommendation performance declines across all baselines, which is an expected outcome. 
However, in the context of the unlearning task, our primary concern is the ability to preserve recommendation performance on the remaining samples rather than the unlearning effectiveness alone.

From the table, 
We observe that the recommendation performance of SISA and SRU\footnote{
SRU results appear lower as we strictly evaluate next-item prediction on the final interaction of each session, whereas the original SRU uses a less restrictive global-sequence average across all sub-sequences.
} drops markedly, possibly due to the partitioning of the dataset into disjoint shards, which substantially reduces correlations between sessions.
Even with the incorporation of attention-based aggregation mechanisms in SRU, its effectiveness in mitigating performance degradation remains limited.
Compared with Retrain, their performance drops by 63\% and 32\%, respectively. 
Such a drastic reduction is undesirable, as overly low recommendation performance would directly impair the user experience.
 
In contrast, CAU demonstrates significantly better performance preservation, achieving results close to the Original model with a stable performance drop controlled within 8\%. 
This advantage can be attributed to the introduction of the \textit{normal loss} and \textit{KL loss} during the unlearning model update, which suppresses excessive disturbance to the original representation distribution, thereby enabling effective unlearning of target samples while retaining the model’s recommendation capability as much as possible.

\subsection{Unlearning effectiveness and efficiency}
\subsubsection{Unlearning effectiveness.}
The results of unlearning performance are summarized in Table \ref{table:unlearning_effectiveness}. 
The Original model represents the effect without unlearning, showing very high $Hit_u$ values. 
Retrain is the most ideal yet computationally expensive approach,
but due to the abundant correlations and connections inherent in session-based recommendation, its $Hit_u$ value is not always the lowest. 
However, the recommendation performance of SISA and SRU drops significantly. 
In this case, even if the $Hit_u$ metric is low, it holds little practical significance.
To enhance comparability among models, the U-score is employed to jointly measure both recommendation performance and unlearning effectiveness.
In the overall comparison, Retrain achieves a relatively high U-score as it attains a well-balanced trade-off between recommendation performance and unlearning effectiveness. 
Although the Original model exhibits strong recommendation performance, it performs poorly in unlearning, resulting in a low U-score. 
Both SISA and SRU suffer from unstable and suboptimal U-scores due to their inability to maintain adequate recommendation quality. 
But CAU consistently achieves the best performance, maintaining recommendation accuracy close to Original while attaining a lower $Hit_u$ score than Retrain.

Importantly, we can observe that in most cases, soft-sampling is more effective than hard-sampling. 
For instance, on the ML-1M and BERT4Rec datasets using gradient unlearning difficulty, the U-score of soft-sampling is 0.6913 compared to 0.6711 for hard-sampling. 
Similarly, with embedding unlearning difficulty, soft-sampling consistently achieves higher scores than hard-sampling.
This improvement is likely due to the increased robustness that soft-sampling provides throughout the unlearning procedure.
\begin{table}[htbp]
\centering
\caption{Comparison of unlearning efficiency (minutes, m). Boldface denotes the best score.}
\begin{tabular}{l cccc}
\toprule
{Method} & {Retrain} & {SISA} & {SRU} & \textbf{CAU} \\
\midrule
GRU4Rec & 30.72m & 3.46m & 5.45m &  \textbf{0.57m}\\
SASRec & 29.21m & 4.54m & 9.04m & \textbf{2.24m} \\
BERT4Rec & 30.24m & 4.64m & 8.47m & \textbf{2.24m} \\
\bottomrule
\end{tabular}
\label{tab:runtime}
\vspace{-10pt}
\end{table}

\subsubsection{Unlearning efficiency.} Table \ref{tab:runtime} compares the unlearning time of the three models on the Games dataset.
It can be observed that the Retrain method is undoubtedly the most time-consuming. 
For SRU and SISA, exact unlearning requires retraining sub-models or aggregation layers, resulting in substantial time overhead.
In contrast, CAU achieves the best overall efficiency by training exclusively on the unlearning data while attaining satisfactory unlearning performance within no more than 200 epochs.
In most cases, CAU achieves a speedup of approximately $13\times$ over Retrain and $3\times$ over SISA and SRU, which further demonstrates the  efficiency of our approximate unlearning method.

\begin{figure*}[htbp]
  \centering
  \includegraphics[width=0.9\textwidth]{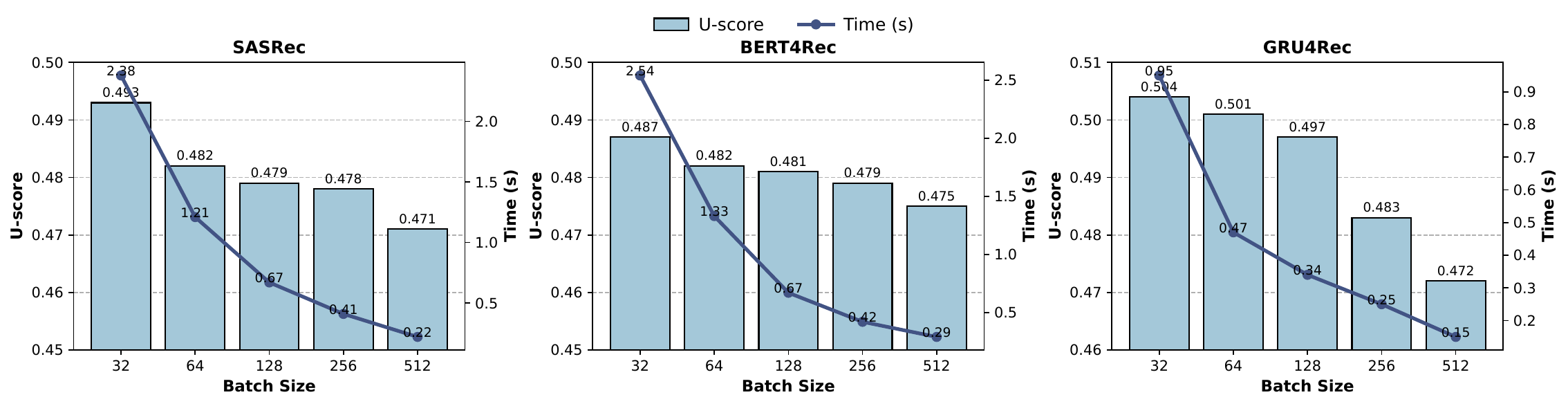}
  \caption{Impact of the unlearning batch size.}
  \label{fig:batch_size}
\end{figure*}

\begin{table}[htbp]
\centering
\caption{Ablation study on three datasets with SASRec as backbone model. Boldface denotes the best scores. w/o denotes without. N is short for NDCG, R is short for Recall.}
\resizebox{0.9\columnwidth}{!}{
\begin{tabular}{llcccc}
\toprule
Dataset & Variant & N@10$\uparrow$ & R@10$\uparrow$ & Hit$_u$@1$\downarrow$ & U-score$\uparrow$ \\
\midrule
\multirow{4}{*}{Beauty} 
& w/o PS & 0.0236 & 0.0503 & 0.0976 & 0.7727 \\
& w/o CU & 0.0254 & 0.0499 & 0.1106 & 0.7624 \\
& w/o RL & 0.0010 & 0.0027 & \textbf{0.0165} & 0.2127 \\
\cmidrule(lr){2-6}
& \textbf{CAU} & \textbf{0.0255} & \textbf{0.0522} & 0.0848 & \textbf{0.7865} \\
\midrule
\multirow{4}{*}{Games} 
& w/o PS & 0.0453 & 0.0845 & 0.0486 & 0.4695 \\
& w/o CU & 0.0480 & 0.0933 & 0.1572 & 0.4673 \\
& w/o RL & 0.0009 & 0.0027 & \textbf{0.0030} & 0.0256 \\
\cmidrule(lr){2-6}
& \textbf{CAU} & \textbf{0.0502} & \textbf{0.0974} & 0.1402 & \textbf{0.4823} \\
\midrule
\multirow{4}{*}{ML-1M} 
& w/o PS & 0.0769 & 0.1526 & 0.0269 & 0.6329 \\
& w/o CU & 0.1021 & 0.1966 & 0.1491 & 0.6384 \\
& w/o RL & 0.0012 & 0.0031 & \textbf{0.0041} & 0.0302 \\
\cmidrule(lr){2-6}
& \textbf{CAU} & \textbf{0.1129} & \textbf{0.2132} & 0.1366 & \textbf{0.6616} \\

\bottomrule
\end{tabular}
\label{tab:ablationstudy}
}
\vspace{-4pt}
\end{table}


\subsection{Ablation study}


In this section, we conduct an ablation study to analyze the functionality of the main components and strategies of CAU (i.e., Pareto-optimality solver (PS), curriculum unlearning (CU), and recommendation loss (RL), where RL consists of both the normal loss and the KL loss). 
Table \ref{tab:ablationstudy} shows the performance of CAU and its variants on all three datasets using SASRec as the backbone model. 
Next, we introduce the variants and analyze the effects.

(1) Effect of the Pareto-optimality solver:
We removed the Pareto-optimality solver component and assigned an equal weight to each task, which means setting all $\alpha$ values to 0.33.
The experimental results demonstrate the effectiveness of the Pareto-optimality solver: by formulating the entire unlearning process as a multi-task optimization problem, the Pareto-optimality solver iteratively searches for gradient update directions that do not degrade any individual sub-task, thereby achieving a better balance between recommendation performance and unlearning effectiveness.

(2) Effect of curriculum unlearning:
We remove the curriculum unlearning strategy and unlearn the data randomly in each epoch.
This proves the effectiveness of our proposed method: guiding the model to unlearn data from easy to difficult based on their unlearning difficulty scores enables it to achieve a satisfactory forgetting level more quickly.

(3) Effect of recommendation loss:
We remove both the \textit{normal loss} and the \textit{KL loss}, which means directly apply GA to unlearn data.
We see that this will lead to catastrophic effects, as the model’s recommendation accuracy almost completely collapses within fewer than five unlearning epochs, and the $Hit_u$ value becomes meaningless. 
This observation highlights the critical role of the recommendation loss in preserving the model’s recommendation capability.

\subsection{Hyperparameter study}
\subsubsection{Impact of unlearning batch size}
Next, we investigate the impact of batch size on the unlearning performance during the forgetting process.
Because the Pareto-Optimality solver estimates task conflicts and updates the model on a batch basis, the batch size plays a critical role in determining its effectiveness.

Here we choose batch size $\in \{32,64,128,256,512\}$ to conduct experiments. 
Figure~\ref{fig:batch_size} illustrates the performance of the three models on the Games dataset.
We clearly observe that as the batch size increases, the U-score tends to decrease. 
This is because the Pareto-optimality solver dynamically searches for the optimal gradient update on a per-batch basis.
And a smaller batch size reduces conflicts among data samples within each batch, increasing the likelihood of identifying accurate update directions, thereby yielding higher U-scores, but may lead to longer training time. 
We also measure the training time per epoch under different batch sizes and find that training time decreases as the unlearning batch size grows. 
For instance, on BERT4Rec, the training time per epoch is 0.4 seconds with a batch size of 256, whereas it is 2.5 seconds with a batch size of 32. 
The model generally converges to satisfactory unlearning performance after approximately 100 epochs. 
This demonstrates a trade-off between unlearning effectiveness and efficiency when selecting the unlearning batch size.

\subsubsection{Impact of unlearning data ratio}
\begin{figure}[htbp]
  \centering
  \includegraphics[width=0.9\columnwidth]{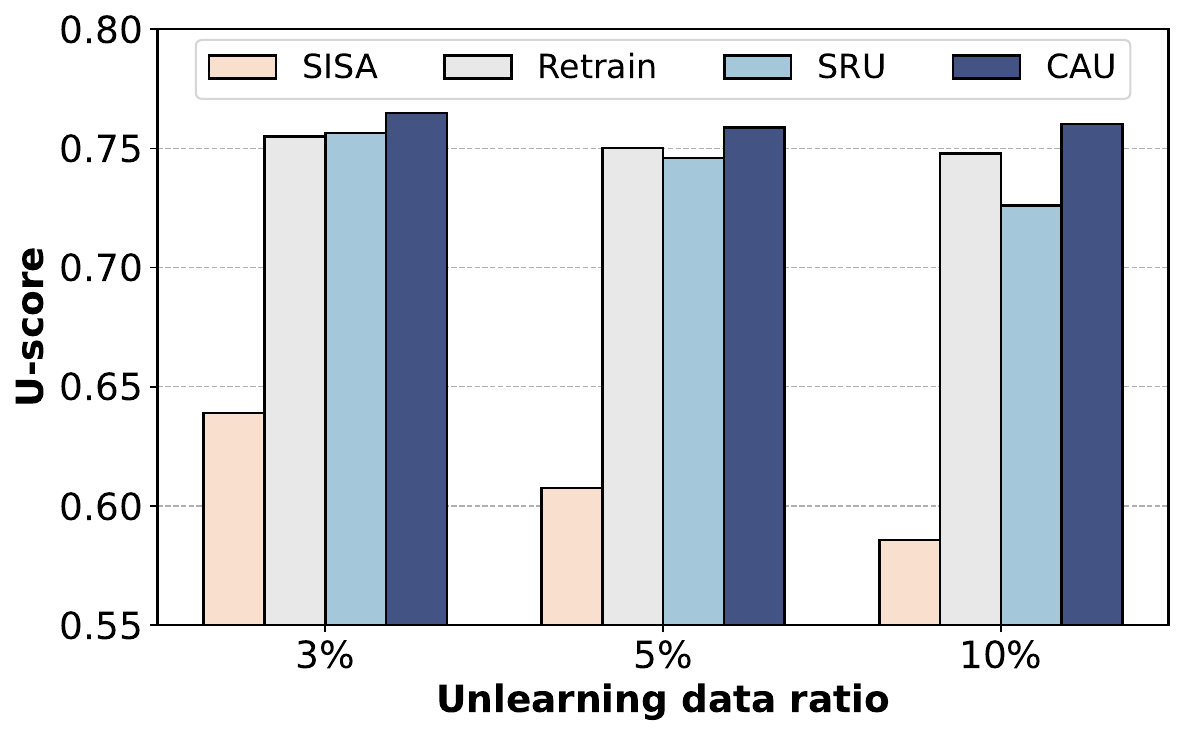}
  \caption{Impact of the unlearning data ratio.}
  \label{fig:unlearnratio}
\end{figure}

Finally, we investigate the performance of different models under varying amounts of unlearning data, as illustrated in Figure~\ref{fig:unlearnratio}.
GRU4Rec is chosen as the backbone recommender model and trained on the Beauty dataset.
We set the unlearning data ratio of the training data from 3\% to 10\%.
We can observe that the variation in U-score for Retrain is minimal, as it represents the most ideal yet computationally expensive approach. 
CAU also exhibits a variation within 1\% when the ratio changes, further indicating the greater stability and robustness of our method, particularly due to the contributions of the Pareto-optimality solver and sampling strategy. 
In contrast, the stability of SISA and SRU decreases as the amount of unlearning data increases, with SISA dropping by 8\% and SRU by 4\%.

\section{Conclusion}
In this paper, we have proposed CAU, a curriculum-based approximate unlearning framework specifically designed for session-based recommendation systems. 
By formulating unlearning as a multi-objective optimization problem and seeking Pareto-Optimal solutions, CAU effectively balances the trade-off between forgetting targeted samples and preserving overall recommendation performance. 
Furthermore, the curriculum learning strategy, guided by the proposed gradient and embedding unlearning difficulty scores, along with the hard-sampling and soft-sampling methods, enables a more efficient and effective unlearning process.
Extensive experiments on multiple benchmark datasets demonstrate the superiority of CAU over existing methods, confirming its practical value and robustness in handling multiple unlearning requests. 

Nevertheless, certain limitations still persist in the current study. 
While our curriculum strategy effectively prioritizes unlearning batches, the long-term impact on model fairness across heterogeneous session lengths remains understudied. 
Regarding potential directions for future work, we plan to further validate CAU's generalizability by adapting it to diverse recommendation architectures beyond session-based recommendation, such as graph-based or cross-domain recommendation models.

\clearpage
\bibliographystyle{ACM-Reference-Format}
\balance
\bibliography{references}

\appendix

\end{document}